# A Jet-Stirred Apparatus for Turbulent Combustion Experiments

Abbasali A. Davani; Paul D. Ronney
University of Southern California
Los Angeles, California, United States

A novel jet-stirred combustion chamber is designed to study turbulent premixed flames. In the new approach, multiple impinging turbulent jets are used to stir the mixture. It is well known that pair of counterflowing turbulent jets produces nearly a constant intensity $(u')$ along the jet axes. In this study, different numbers of impinging jets in various configurations are used to produce isotropic turbulence intensity. FLUENT simulations have been conducted to assess the viability of the proposed chamber. In order to be able to compare different configurations, three different non dimensional indices are introduces. Mean flow index; Homogeneity index, and Isotropicity index. Using these indices one can compare various chambers including conventional Fan-stirred Reactor. Results show that a concentric inlet/outlet chamber with 8 inlets and 8 outlets with inlet velocity of 20 m/s and initial intensity of 15% produces near zero mean flow and 2.5 m/s turbulence intensity which is much more higher than reported values for Fan-stirred chamber.

## 1  Introduction

It is well known that turbulence increases mean flame propagation rate $S_T$ ($\frac{S_T}{S_L} = (10)$) and thus mass burning rate (= $\rho S_T A_{Projected}$). Therefore, higher heat produced per unit volume per unit time which means higher thermal efficiencies. However, as it can be seen in Figure 1 neither theoretical models nor computational studies have yet been able to explain these observations. In fact, the theoretical models do not agree with the experiments nor with each other.

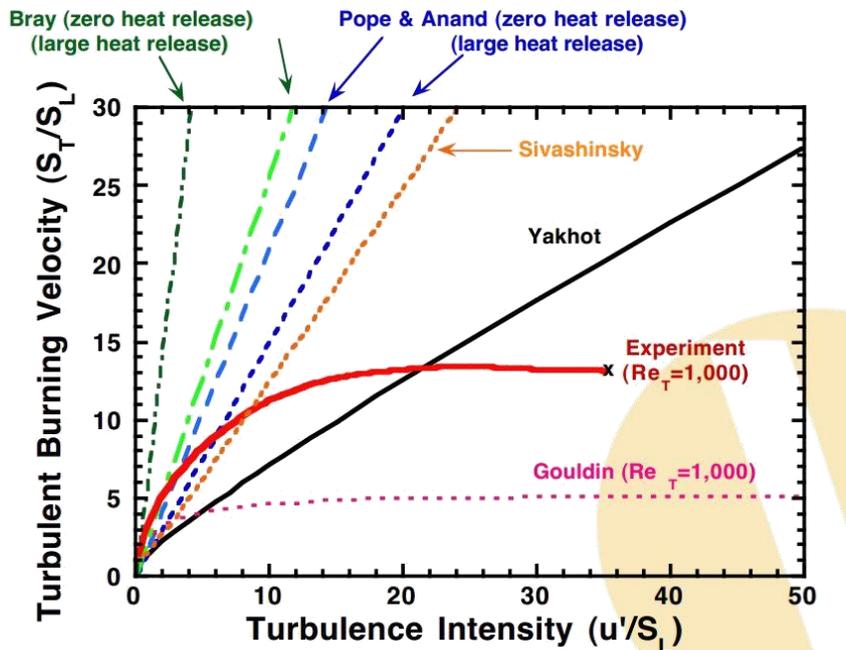

Figure 1. Comparison between various theoretical and computational studies.

In contrast to the existing models and computation, experiments show that mixing faster will not causes faster burning rate and it could cause extinguishment of the mixture. We believe that this phenomena could be caused by the experiment itself i.e. the apparatus that is used to study turbulent flames which is using fans to generate turbulence intensity.

A novel jet-stirred combustion chamber is designed to study turbulent premixed flames. In the new approach, multiple impinging turbulent jets are used to stir the mixture. It is well known (Geyer *et al.*, 2005; Mastorakos *et al.*, 1995; Sardi *et al.*, 1998; Kempf *et al.*, 2000) that pair of counterflowing turbulent jets produces nearly a constant intensity $(u')$ along the jet axes. In this study, different numbers of impinging jets in various configurations are used to produce isotropic turbulence intensity.

Compared to the traditional fan-stirred chamber, there are several advantages:

(1) Any number of jets can be used to create a nearly isotropic flow.
(2) A single pump, external to the combustion chamber, can be used to power the flow.
(3) There are no shafts penetrating the chamber wall that need to be sealed, only static jets and ports.
(4) There is no flow bias due to the swirl created by fans.
(5) Any desired amount of swirl can be introduced in a well-controlled manner.

In addition, jet-stirred chambers retain most of the advantages of fan-stirred chambers, i.e.
(1) $u'$ is independent of the mean flow (unlike most other flows such as jet or grid turbulence) and can be made very large compared to $S_L$.
(2) The walls are remote, thus conductive heat losses are negligible.
(3) The flames are free to propagate at any rate they choose.
(4) $S_T$ is easily measured.
(5) The flames are not subject to a mean strain (as in a counterflow) or mean shear.
(6) The effects of pressure are readily assessed.

## 2 Numerical Model and Validation

FLUENT simulations have been conducted to assess the viability of the proposed chamber. However, before starting the computations several simulations are carried out using existing experimental data and computations and results are compared in order to ensure the validity of used numerical models. Pettit, M.W.A. *et. al.* (2010) measured axial and radial velocity fluctuations ($u'$ and $v'$) between a pair of impinging jets experimentally and performed a Large-Eddy simulation for the comparison. The experiment is simulated in FLUENT and results are compared in Figure 2.

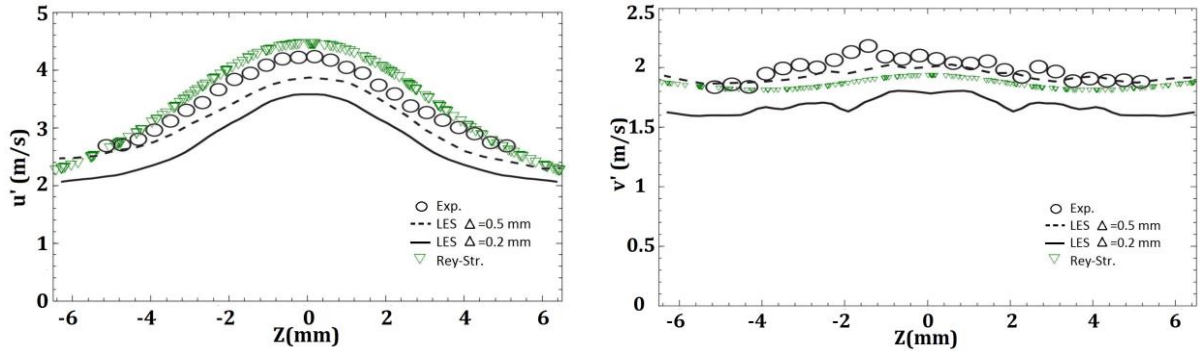

Figure 2. Axial (u') and radial (v') fluctuations (m/s) comparison. Green triangle is FLUENT simulation.

As it can be seen, results are in a good agreement with both experiment and Larg-Eddy simulation.

Ravi S. *et al.* (2012) in an experimental study of turbulent statistics in a fan-stirred reactor reported the maximum turbulence intensity and mean flow measurements. The experiment is simulated in FLUENT and results are presented in Table 1.

| Results | Mean U (m/s) | Maximum $u'$ (m/s) |
|---|---|---|
| Experiment | 0.3 | 1.48 |
| Simulation | 0.26 | 1.45 |

Table. 1. Mean flow and intensity comparison between experiment and simulation.

In this case also results from the simulation are in agreement with the experimental measurements.

## 3 Apparatus Simulation

Once the validity of the models used are ensured, proposed jet-stirred reactors with various numbers of jets are simulated. Jet numbers varies from 4 to 92 and different configurations based on platonic solids are used. Contours of mean flow and turbulence intensity along different axes for concentric jet-stirred chamber based on a tetrahedron are presented below:

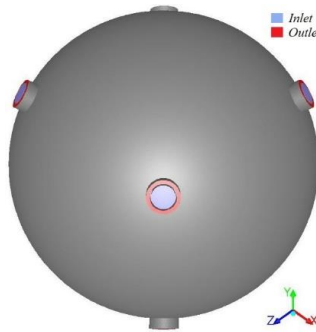

Figure 3. Concentric jet configuration.

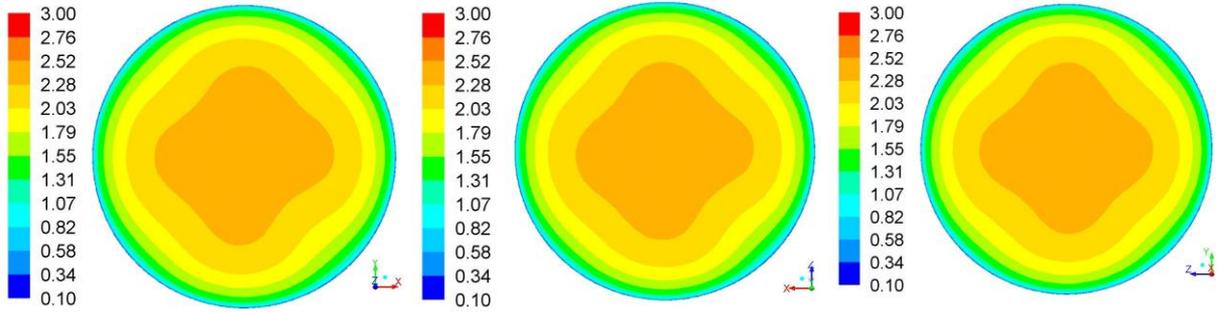

Figure 4. Contours of Turbulence Intensity (m/s) in three different planes for concentric apparatus.

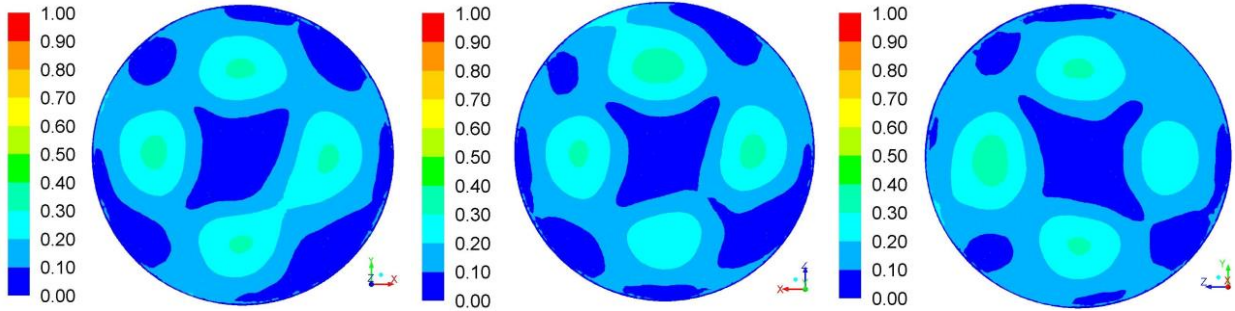

Figure 5. Contours of Velocity Magnitude (m/s) in three different planes for concentric apparatus.

Profiles of intensity and velocity magnitude for concentric apparatus are shown in Figure 6.

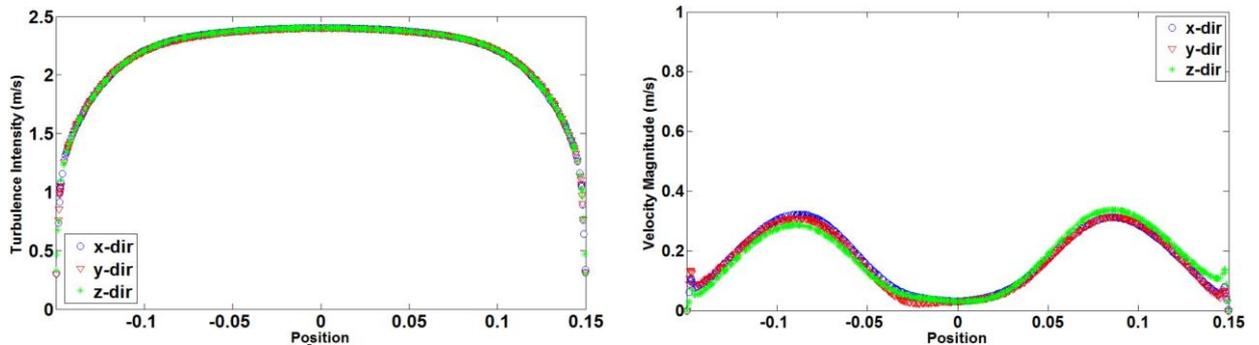

Figure 6. Profiles of Turbulence intensity (m/s) and Velocity (m/s) along three different axes. for concentric apparatus.

## 4    Result and Discussion

As it can be seen, the chamber produces same turbulence intensity in different directions and at the same time the mean flow is near zero. In order to be able to compare different chamber configurations, three different non dimensional indices are introduces. Mean flow index; Homogeneity index, and Isotropicity index. These indices are defined below:

$$V = \sum_1^n v_i \quad ; \qquad MFI = \sqrt{\frac{\sum_{i=1}^n \overline{a_i}^2 v_i}{V}} \Big/ \overline{a'_V}$$

$$\overline{a'_V} = \sqrt{\frac{\sum_{i=1}^n ({u_i'}^2 + {v_i'}^2 + {w_i'}^2) v_i}{3V}} \quad ; \qquad HI = \frac{\sum_{i=1}^n [(u_i' - \overline{a'_V})^2 + (v_i' - \overline{a'_V})^2 + (w_i' - \overline{a'_V})^2] v_i}{3V}$$

$$a'_i = \sqrt{\frac{(u_i'^2 + v_i'^2 + w_i'^2) v_i}{3V}}$$

$$II = \frac{\sum_{i=1}^{n}[(a'_i - u_i')^2 + (a'_i - v_i')^2 + (a'_i - w_i')^2] v_i}{3V}$$

| | Nomenclature |
|---|---|
| $\overline{a_i}$ | Velocity magnitude at $i^{th}$ cell |
| $v_i$ | Volume of $i^{th}$ cell |
| $V$ | Total volume |
| $n$ | Number of cells |
| $u'$ | Turb. intensity comp. in x-dir |
| $v'$ | Turb. intensity comp. in y-dir |
| $w'$ | Turb. intensity comp. in z-dir |
| $\overline{a'_V}$ | Average intensity over V |
| $a'_i$ | Average intensity at each cell |

Homogeneity here means that intensity is statistically independent of the shift of the coordinate system and Isotropicity means that intensity is statistically independent of the rotation of the coordinate system. Using these indices one can compare various chambers including conventional Fan-stirred Reactor. Results show that a concentric inlet/outlet chamber with 8 inlets and 8 outlets with inlet velocity of 20 m/s and initial intensity of 15% produces near zero mean flow and 2.5 m/s turbulence intensity which is much more higher than reported values for Fan-stirred chamber.

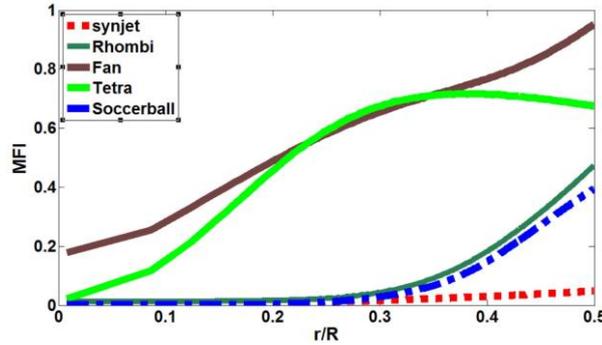

Figure 7. MFI comparison of various jet-stirred chambers.

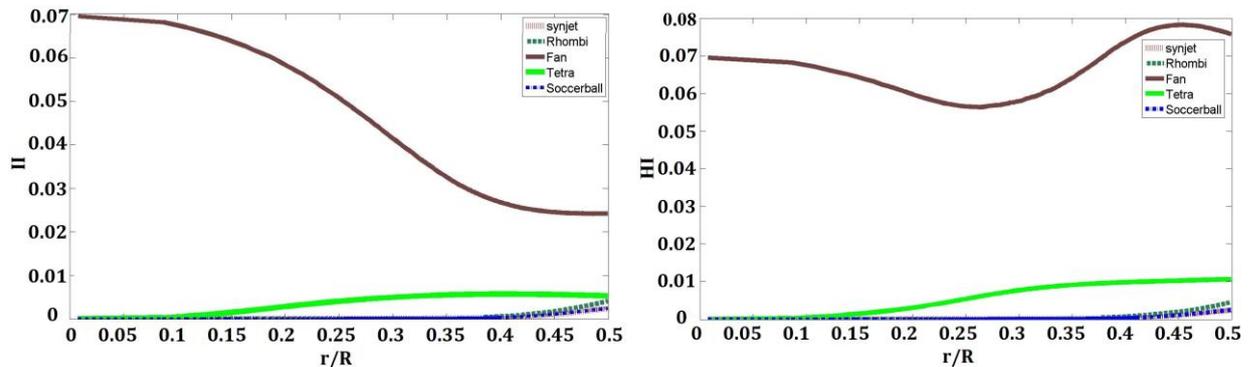

Figure 8. HI and II comparison of various jet-stirred chambers and a fan-stirred reactor.

In the Figures 7-8 the vertical axis is index and horizontal axis shows moving from center toward the wall of chamber. As radius is increased from center of the reactor, the less the index is deviated the better. As it can be seen concentric configuration is the best compared to other in terms of producing isotropic homogeneous intensity with near zero mean flow. Also, jet-stirred reactors in general perform better than fan-stirred reactor. After performing cold flow analysis, a premixed combustion is simulated to study the propagation characteristics of flame. Contours of progress variable between a pair of jets in 3 different time steps are shown below:

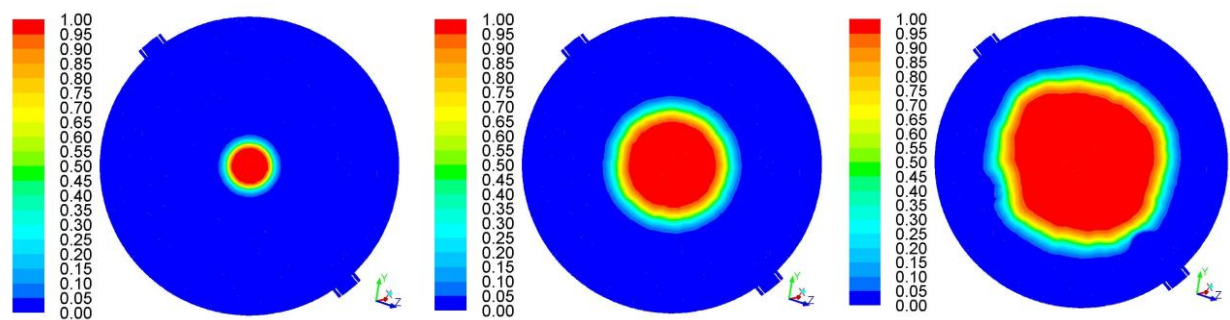

Figure 9. Contours of Progress Variable in three different time steps for concentric apparatus.

It can be seen from Figure 9. That the flame is expanding spherically for more than half of the chamber before it is disturbed by the jets.

Based on the simulations, results show that the novel jet-stirred reactor could be a better apparatus to study turbulent flames since high values of turbulence intensity could be produced with high Isotropicity and homogeneity. Among different reactor modeled, concentric inlet/outlet chamber with 8 inlets and 8 outlets with inlet velocity of 20 m/s and initial intensity of 15% produces near zero mean flow and 2.5 m/s turbulence intensity which is much more higher than reported values for Fan-stirred chamber.